# Gasdynamic Diode: How to Stop 100-kV Streamer

A Yu Starikovskiy[1,*] and N L Aleksandrov[2]

[1]*Princeton University, Princeton, NJ08544, USA*

[2]*Moscow Institute of Physics and Technology, Dolgoprudny, 141700, Russia*

[*]Author to whom any correspondence should be addressed

E-mail: astariko@princeton.edu

**Abstract**

The conditions were found when the gaseous medium demonstrates a unidirectional conductivity on a short time scale; a gas density discontinuity forms a kind of "gas-dynamic diode" that allows the plasma channel to propagate in one direction and blocks its development in another. The results of a two-dimensional numerical simulation of a streamer discharge developing through a shock wave in air were presented for various neutral density discontinuities across the wave. The focus was on the case when the streamer propagated from a low-density region to a high-density region. Streamer characteristics changed greatly after intersecting the shock wave. It was shown that the streamer failed to penetrate into the high-density region when the ratio between the densities in these regions was sufficiently high ($> 1.2$). In this case, the discharge developed along the surface between these regions after reaching the boundary between them. Streamers could penetrate into any of the high-density and low-density regions when a neutral particle density discontinuity was replaced by a gradual density change.

# Contents

## Introduction

Development of streamers and streamer-like discharges in heterogeneous gas media could be observed under natural and laboratory conditions. Lightning discharges propagate in the cloud-to-ground gap along which the air density $n$ changes several times gradually [1, 2]. Lightning-initiated transient luminous phenomena (red sprites, blue jets, elves) develop at high altitudes and cover tens of kilometers where the air density is subjected to wide variations (up to several orders of magnitude) [2, 3]. Experimental modeling of red sprites using a hot jet which imitates a gradient density in air was performed in [4]. This has shown the strong influence of the density variation on streamer propagation. Streamers also interact with strong density discontinuities when the discharges propagate in a gas containing polarized solid particles [5] and water droplets [6, 7]. Streamer-like discharges intersect neutral particles density discontinuity in a shock wave formed in front of supersonic aircraft and spacecraft when lightning discharges develop from them. Over 90% of lightning strokes to these objects are due to the discharges excited by the aircraft and spacecraft themselves [1, 2]. In the initial stage, these discharges are so-called leaders with streamers generated at the leader tips.

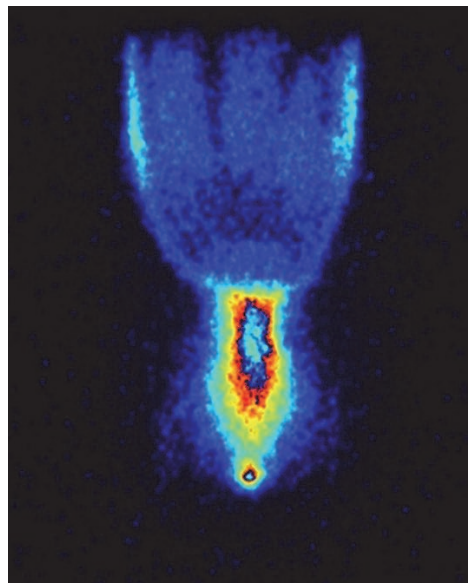

Figure 1. High-voltage streamer development in a complex geometry.
Voltage $U$ = 20 kV, 5 ns after the discharge start. ICCD camera gate is 1 ns.
Air, $P$ = 1 atm, $T$ = 620 K, interelectrode distance $h$ = 10 mm.

Figure 1 shows nanosecond discharge development in the hot air flow. The distance between the sharp high-voltage electrode (bottom) and the rounded ground electrode covered with a quartz dielectric layer (top) is 10 mm. Discharge propagates as a free streamer in the interelectrode gap and develops as a surface wave along the dielectric surface. The velocity of the propagation of the plasma layer reaches 3 mm/ns (~1% of the speed of light).

Interaction between nonequilibrium discharges and high speed flows is important for plasma aerodynamic flow control and is characterized by high gas inhomogeneities produced in shock waves and rarefaction waves. The influence of nonequilibrium weakly ionized plasmas generated in different discharges (DC, RF and microwave) on shock waves was thoroughly studied under laboratory conditions [8], whereas the effects of gas flow inhomogeneities on discharge plasmas has been poorly understood.

Streamer and streamer-like discharges (for example, nanosecond surface dielectric barrier discharges - SDBDs) were widely used in plasma aerodynamics to control airflow by generating weak shock waves due to fast, local gas heating [9-11]. These discharges can efficiently ignite combustible mixtures (see [12-15] and references therein). These mixtures may involve inhomogeneities such as fuel jets and fuel aerosols or droplets. Therefore, studying streamer interaction with shock waves and with some other inhomogeneities is of practical importance. For example, it is necessary to know the spatial distribution of deposited discharge energy and gas heating, as well as the distribution of energy input over various degrees of freedom of molecules.

These effects are difficult to study experimentally because a streamer discharge develops usually in the form of a number of filamentary channels (see, for instance, [1]). At present numerical calculations are widely used to simulate streamer development under various conditions (see, for instance, [16]). In particular, the influence of gas density gradients on streamer propagation was simulated for atmospheric high-altitude discharges [17, 18] as well as for discharges that intersect bubbles and particles [19, 20]. In our previous work [21], a streamer discharge propagating through a shock wave from a high-density region to a low-density one was studied both experimentally and numerically. It was shown that the streamer properties change drastically when intersecting the shock surface.

In the present work we consider the opposite case when a streamer discharge intersects a shock wave and develops from a low-density region to a high-density region. It was shown that streamer behavior can differ significantly from the case when the discharge propagates in the opposite direction. In some cases, the streamer cannot penetrate into the high-density region and the discharge propagation is completely blocked by the gas density discontinuity. In addition, we considered how the streamer develops through inhomogeneities in which the gas density changes gradually (the case of rarefaction and compression waves).

## Numerical Model

To demonstrate the peculiarities of the interaction of a streamer with a shock wave, we simulated a single cathode-directed streamer propagating in a 15-cm discharge gap, in which the gas density $n$ changed as a step function of the distance from the anode. The density was equal to $n_1$ for a distance less than 7 cm from the tip of the high-voltage electrode as well as to $n_2$ for the gap between this point and the grounded cathode. The grounded electrode was a plate, while the high-voltage electrode was a plate with a needle in the center. The needle shape was a semi-ellipsoid with a major semi-axis of 0.8 cm and a minor semi-axis of 0.08 cm. The streamer started from the needle tip. Both the case where the discharge started in the region of high gas density and penetrated into the low-density region ($n_1 > n_2$), and the case where the discharge started in the low-density region and developed in the region of high gas density ($n_1 < n_2$) were considered. The voltage across the gap increased linearly over time (0-1 ns), and then remained constant at the 100 kV level. The geometry of the discharge gap is shown in Figure 2.

The upper part of the figure shows the isolines of the electric field in the gap 15 ns after the start of the high-voltage pulse. The air pressure in the discharge gap is 250 Torr, the maximum voltage is 100 kV. To initiate the start of the streamer near the tip of the high-voltage electrode, a small quasi-neutral preionization region was numerically set. An axially symmetric 2D model based on the hydrodynamic approximation of the electron ensemble motion was used. The model itself, the method of calculation and the verification of the numerical model using experimental data were described in detail in [22-26]. In the model, the transport equations for charged particles (electrons, positive and negative ions) were solved numerically along with the Poisson equation for the electric field on an adaptive grid. The minimum size of the grid cell was 1 μm. The minimum time step was $5\times10^{-14}$ s. The bottom part of Figure 2 illustrates the principles of the adaptive mesh construction. The size of the computational cell decreased gradually in areas of a high electric field and large gradients. It also increased towards the boundaries of the computational domain.

The electron mobility, ionization rate coefficients and attachment rate coefficients were calculated from the Boltzmann equation using the two-term approximation and the local electron energy distribution function (EEDF) approximation. Under the conditions studied plasma was a weakly-ionized one in which the electron transport and rate properties including the ionization coefficients were controlled by the reduced electric field $E/n$; that is, the ionization rate in rarified air was higher than that in dense air, whereas the breakdown electric field increased with increasing air density. The photoionization of gas was described on the basis of the Zheleznyak-Mnatsakanian-Sizykh model [27]. The model provides a quantitative self-consistent, photoionization distribution in air-like mixtures.

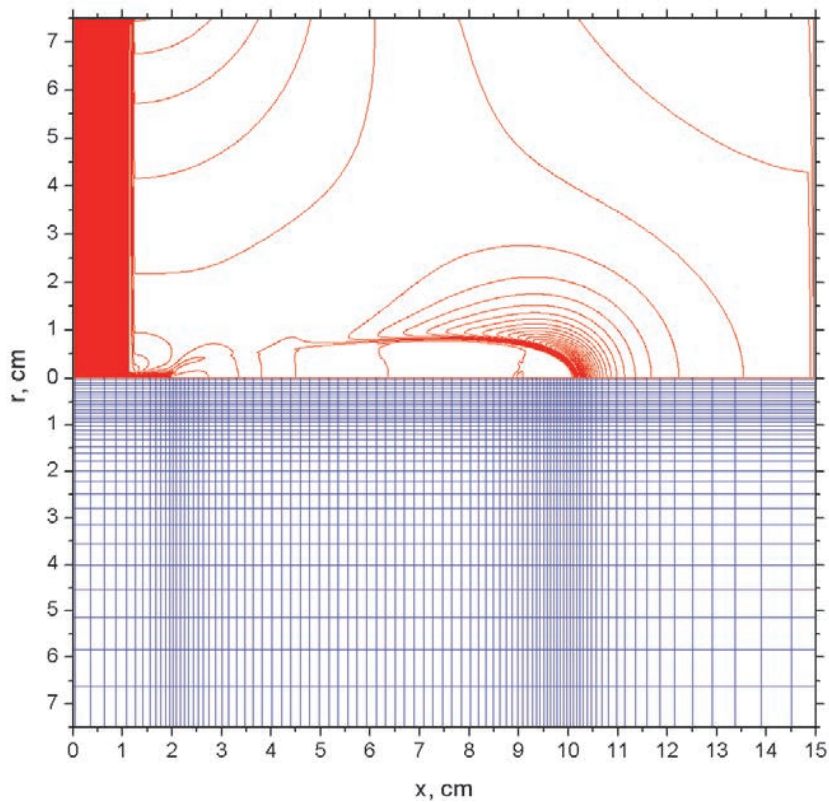

Figure 2. Discharge gap geometry and adaptive computational mesh (every $10^{th}$ cell is shown). The top shows the contours of the electric field of the streamer 15 ns after the start of the high voltage pulse.
$P = 250$ Torr, $T = 293$ K, $U = 100$ kV, air.

## Results

### Streamer propagation through a strong density discontinuity

All calculations were carried out for the same initial gas pressure $P = 250$ Torr, gas temperature $T = 300$ K and mixture composition (synthetic air $N_2:O_2 = 80:20$), with a voltage between the electrodes of 100 kV. Figure 3 demonstrates the dynamics of a streamer discharge development in the presence of a gas density discontinuity in the gap. At the point $x = 7$ cm, the concentration of the gas abruptly decreases by 2.07 times. In the first two images of Figure 3 (8 and 10 ns from the start) the streamer discharge develops as usual; here, a conducting channel with a diameter of about 10 mm is formed at a speed of approximately 7-8 mm/ns. When the streamer head approaches the plane of the density discontinuity (which can be, for example, a shock wave moving in the same direction as the streamer), the picture changes.

The electric field at a large distance in front of the streamer head is not strong enough to ionize a gas at an initial gas density corresponding to a pressure of 250 Torr at normal temperature. The decrease in density by 2.07 times leads to the fact that the electric field in the low-density region becomes above the breakdown threshold. Photoionization forms the seed electrons, and at some distance from the streamer head a plasma region begins to form. Due to the movement of electrons towards the anode (to the left in Figure 3) a significant negative charge accumulates near the density discontinuity plane and charge separation takes place. The ionization wave forming in a rarefied gas carries a significant positive charge at the front (image 10.2 ns in Figure 3).

The presence of a negative charge at the density discontinuity greatly enhances the field in the gap between the streamer and the discontinuity plane and increases the velocity of the ionization wave on the axis. The streamer is deformed. The hemispherical head becomes almost cone-shaped and touches the region of conducting plasma in the low-density region at almost one point (image 10.3 ns). At this moment, the negative charge accumulated at the density discontinuity is neutralized, and a secondary ionization wave begins to propagate to the ionized gas region to the right of this boundary, pushing the potential of the streamer head forward (image 10.4 ns). At the same time, the expansion of the streamer channel in the low-density region begins (image 10.5 ns).

Near the discontinuity boundary, the density of photoelectrons is at its maximum since photoionization in this region occurs due to photons produced in the region of high gas densities and, accordingly, high electric fields at the ionization wave front. Due to the rapid radial expansion of the channel along this pre-ionized region, a characteristic "skirt" is formed at the density discontinuity boundary (image 10.6 ns). The radial component of the field on the right side of this region becomes insignificant due to the shielding by a wide conducting region in the zone of low-density gas. At the same time, the radial expansion of the channel stops, and a characteristic waist is maintained at the streamer channel (image 10.7 ns).

The channel region near the discontinuity surface must conduct the full current of the streamer, because of which the current density increases and the longitudinal field providing the streamer current is greater in this region than in the others (10.8 ns image). The transition of the streamer to a quasi-stationary propagation mode in a low-density environment is completed by the formation of a larger diameter head (image 11.0 ns). The speed of the streamer increases significantly; it now spreads at a speed of about 2 cm/ns (images 11.5-12.0 ns). The electric field at the head is significantly reduced, and the streamer switches to the mode of quasi-stationary motion in a low-density medium (12.5 ns image). The approach of the streamer to the low-voltage electrode (image 13.5 ns) leads to an increase in the electric field near the head, acceleration of the streamer and an increase in its diameter.

Thus, even with a significant discontinuity in the gas density, the streamer easily passes the interface if it moves in the direction of decreasing density.

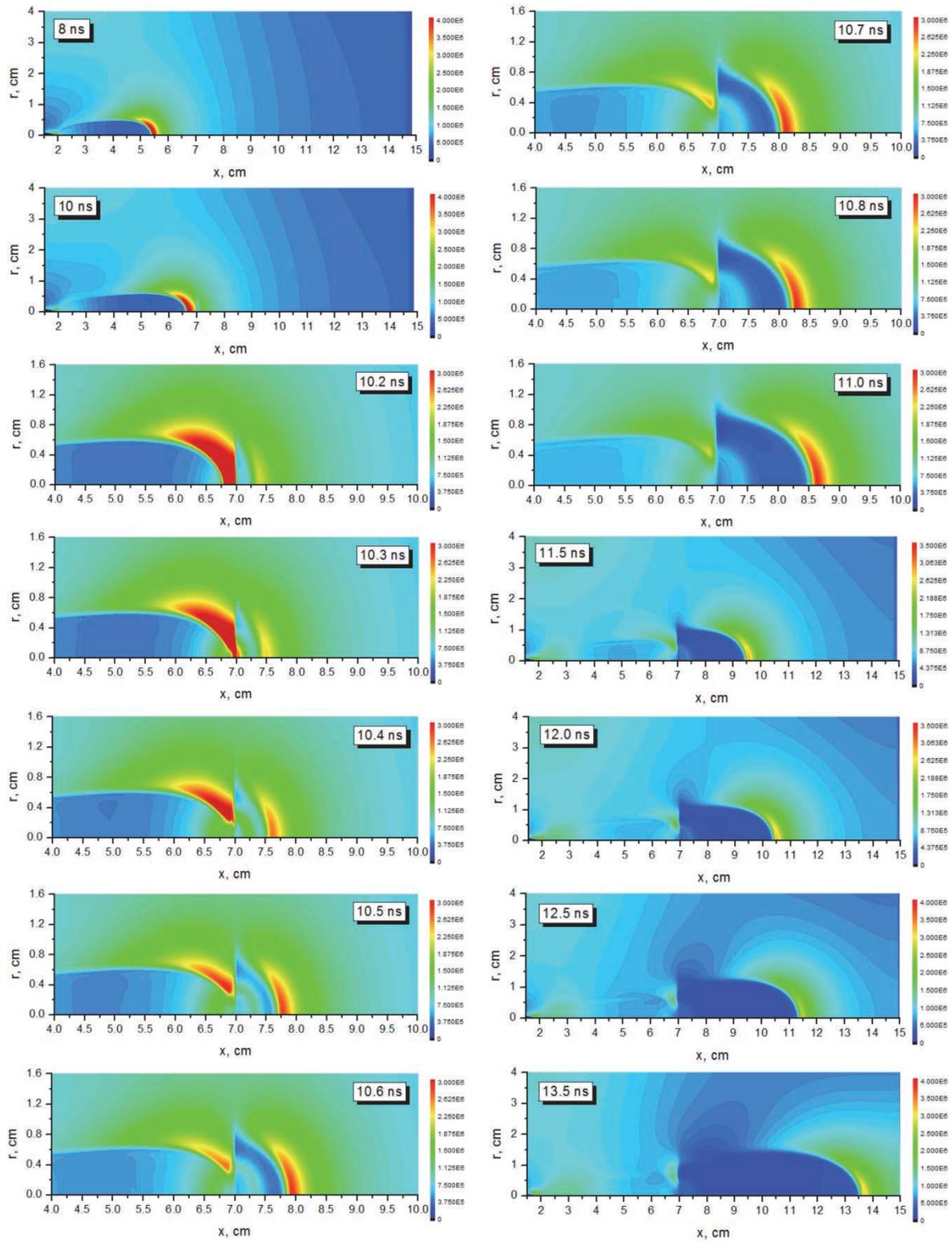

Figure 3. Contours of electric field (color scale is [V/m]) for a streamer propagating from high- to low-density medium. $n_2/n_1 = 0.483$. $P_1 = 250$ Torr. $U = 100$ kV.

A streamer propagating from a low-density region to a high-density region demonstrates different dynamics (Figure 4). In the initial stage, when the streamer moves in a gas with density $n_1 = 8.9 \times 10^{18}$ molec/cm$^3$ (corresponding to an initial pressure of 250 Torr at normal temperature), the development of the process does not differ from the case just considered (Figure 4; images 8 ns and 10 ns). However, when approaching the density discontinuity, the pattern of the discharge development changes. Now the field in front of the streamer head is not enough to ionize the gas beyond the interface, where the gas density is two times higher than initially ($n_2 = 1.8 \times 10^{19}$ molec/cm$^3$). Therefore, until the streamer head arrives at the point of discontinuity, the streamer moves with almost no change in shape (image 10.5 ns).

For further advancement into the high-density gas region, the field formed by a large-diameter head becomes insufficient, and the motion of the ionization wave near the streamer axis slows down dramatically (image 10.6 ns). In this case, the ionization wave is flattened near the axis of symmetry of the channel (image 10.7 ns) and the region of maximum field moves from the axis to the periphery (image 10.8 ns). The highest ionization rate of the gas is shifted to the same point. Furthermore, at this point, the dominant amount of VUV photons responsible for the pre-ionization of the gas ahead of the ionization wave front are produced (image 11.0 ns). As a result, the ionization wave begins to propagate along the gas density discontinuity surface, similarly to the propagation along the surface of the dielectric in the case of a barrier discharge (Figure 1, for example). The principal difference lies in the fact that in this case the role of a "dielectric" is played by the same gas in which the discharge propagates; the only difference is the increased density (image 11.2 ns). The subsequent development of the discharge along the surface leads to the front part of the head becoming flat, and a "plate" of ionized gas with a very high electric field at its outer edge (image 11.5 ns). Further development of the discharge takes place along the surface of the density discontinuity. The diameter of the "plate" increases with a speed of a few millimeters per nanosecond (images 12-14 ns). Moreover, the field on the discharge axis drops, and cannot provoke the development of instability on the axis nor the appearance of a new streamer propagating through the high-density gas (image 16-18 ns). In fact, a discontinuity in the density of a gas with an intensity of 2 is an insurmountable barrier to a streamer discharge.

It becomes extremely interesting to trace the ratio at which the density discontinuity stops the streamer penetration into the medium with a higher density. Figure 5 shows the results of the calculation of streamer propagation in a medium with different density discontinuities at the same time moment. The right column demonstrates the transition of a streamer from a less dense gas to a more dense one (images show time $t = 19.6$ ns after the discharge start), and the left column shows a transition from a more dense to a less dense medium ($t = 14$ ns). The difference in the streamer behavior at different intensities and the direction of the density discontinuity is clearly visible (Figure 5). Even with minimal disturbance (less than 5% for the second row of Figure 5), both in the case of a decrease and in the case of an increase in the density of the gas, the streamer channel shows noticeable distortions. In the case of the streamer passing into a high-density medium, a local channel thickening is formed in the discontinuity plane. In the case of passing into a lower density medium, a local thinning of the channel and an area of increased electric field are observed.

An increase in the intensity of the discontinuity to 9% leads to severe deformations of the streamer channel (third row of Figure 5). There are both a significant change in the channel diameter and a change in the electric fields near the discontinuity. The discontinuity of gas density with an intensity of 20% (fourth row) is the limit for the streamer propagation to a region of high gas density. The secondary off-axis field maximum emerges and leads to the formation of a conducting "plate" in the discontinuity plane. The streamer propagation along the axis decelerates. Furthermore, the instability develops near the discharge axis which leads to the formation of a new head of the smaller diameter streamer, which continues to propagate in a high-density gas.

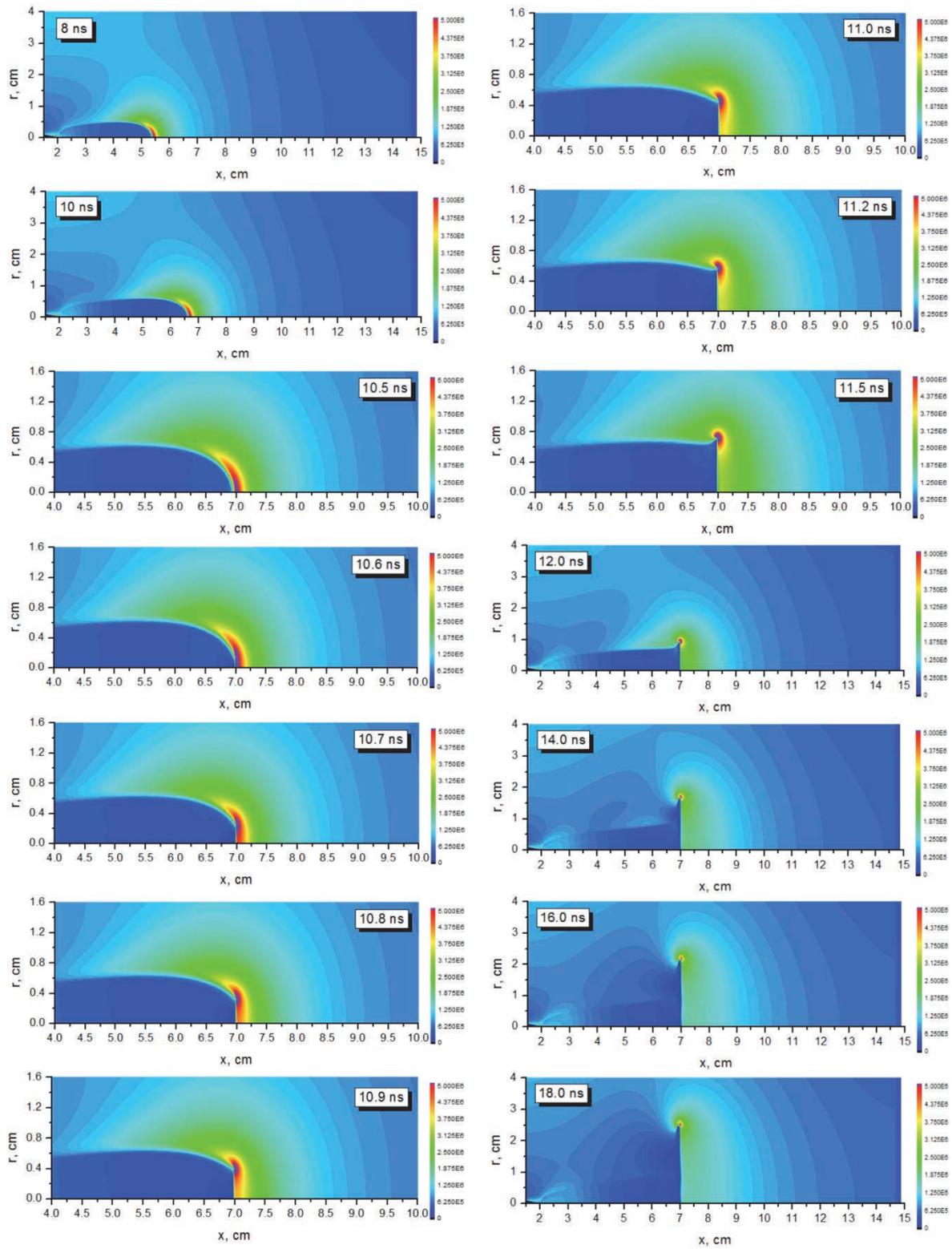

Figure 4. Contours of electric field for a streamer propagating from low- to high-density medium. $n_2/n_1 = 2.07$. $P_1 = 250$ Torr. $U = 100$ kV.

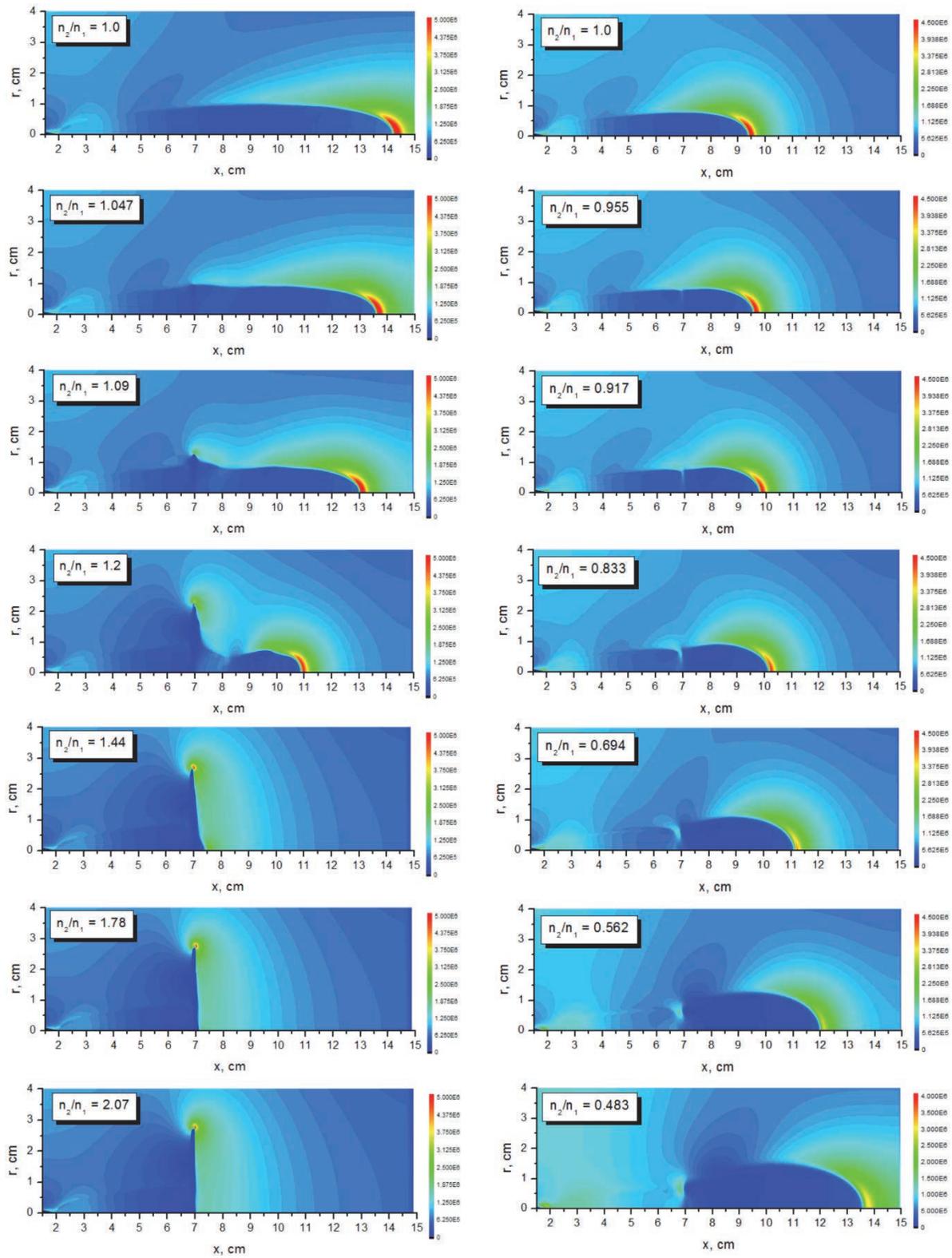

Figure 5. Contours of electric field for a streamer interacting with a discontinuity for various densities ratios. $t = 19.6$ ns (left column); $t = 14.0$ ns (right column).

A further increase in the intensity of the discontinuity leads to an increase in the off-axis maximum of the electric field, and the rapid formation of a plasma layer at the boundary of the density discontinuity. In this case, the field at the discharge axis significantly decreases, and the instability, which could lead to the formation of a streamer in a higher density medium, develops very slowly (Figure 5; row 5). Practically speaking, the streamer cannot penetrate into the region of high-density gas at such a high level of intensity of discontinuity. An increase in the ratio $n_2/n_1 > 1.7$ leads to further enhancement of the effect; the streamer interacts with this discontinuity similarly to a dielectric layer in the case of a barrier discharge.

A decrease in gas density ($n_2/n_1 < 1$), as expected, leads to an increase in the diameter and noticeable acceleration of the streamer (Figure 5; rows 6, 7). In this case, the "waist" on the streamer channel becomes more noticeable in the region of transition to low-density gas, which is associated with the formation of a secondary ionization wave and mutual neutralization of charges on the surface of the discontinuity.

In this case, one can say that the medium for short times began to have unidirectional conductivity; that is, a discontinuity in the gas density formed a kind of "gas-dynamic diode" that allowed the plasma channel to propagate in one direction and blocked its development in the other. Obviously, such a "diode" is not able to operate for longer times. Due to the development of three-dimensional instability in the plasma layer forming the conductive "plate" the symmetry of the electric field is broken, and the discharge begins to propagate to a region of high-density as a new streamer of smaller diameter. However, the development of this instability is a relatively slow process, and for a short period of time the discharge gap with a discontinuity in the gas density can indeed work as a medium with unidirectional conductivity.

Note that a density discontinuity is not the only possible mechanism for the formation of such a directionally conducting medium. A sharp decrease in the frequency of gas ionization rate of any nature could form the same effect (see discussion below). Thus, instead of the density discontinuity, an interface between gases with different ionization cross-sections can be used to create a similar effect (a pair Ar-He, for example).

## Streamer propagation in a near-critical regime

Except for two limiting cases (the absence of a density discontinuity and a very strong discontinuity), the $n_2/n_1 = 1.2$ regime is of considerable interest. This regime is interesting as a threshold between two different cases. For a lower $n_2/n_1$ ratio the streamer is able to pass beyond the discontinuity, and for a higher $n_2/n_1$ it does not. In a near-critical regime like this, the development of a streamer differs qualitatively from the cases of both a large and a small density difference (Figure 6).

The first stage of discharge development, when the streamer is entirely in the region with a constant gas density, does not differ from the cases already considered (Figure 6; images 8 and 10 ns). Then, similarly to the case considered above, with $n_2/n_1 = 2.07$, the ionization wave slows down at the channel axis, and the off-axis electric field maximum is formed (10.5-10.8 ns). However, after this point, the development of the discharge follows a noticeably different scenario.

On the discharge axis, the ionization wave continues to propagate slowly to the region of higher gas density, and the wave front near the discharge axis becomes non-planar (Figure 6; 10.9 ns). Instability begins to develop which leads to a further increase in the convexity of the ionization wave on the discharge axis (Figure 6; 11.0 ns). Further propagation of the ionization wave through high-density gas leads to a sharp decrease in the radius of curvature of the conducting channel head, and to an increase in the electric field on the axis (Figure 6; 11.5 ns). The development of this instability occurs rather slowly (Figure 6; 12.0 ns), but after a while it develops into a streamer channel with a smaller radius, which continues to spread through a high-density gas (Figure 6; 14.0 ns). The developing head of the new streamer gradually leaves the region of low electrical field created by the expanding conductive "plate" (Figure 6; 16.0 ns) and enters the quasi-stationary propagation mode (Figure 6; 18.0 ns). In this case, the discharge retains two field maxima. The first maximum is connected with the edge of the conducting "plate" expanding along the density discontinuity surface, whereas the second one is located on the axis of the newly formed streamer and ensures the discharge propagation in the high-density region.

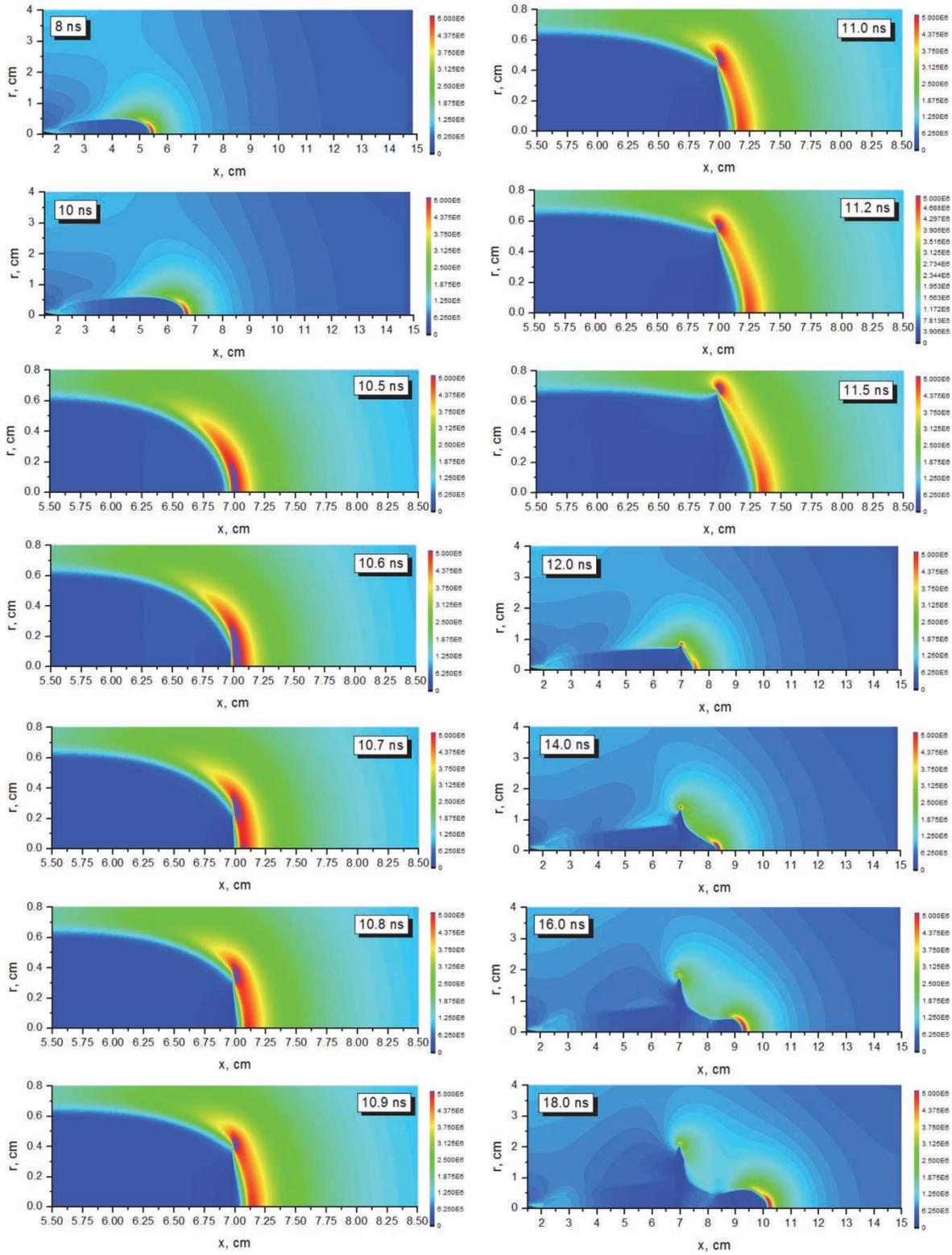

Figure 6. Contours of electric field for a streamer propagating from low- to high-density medium. $n_2/n_1 = 1.20$. $P_1 = 250$ Torr. $U = 100$ kV.

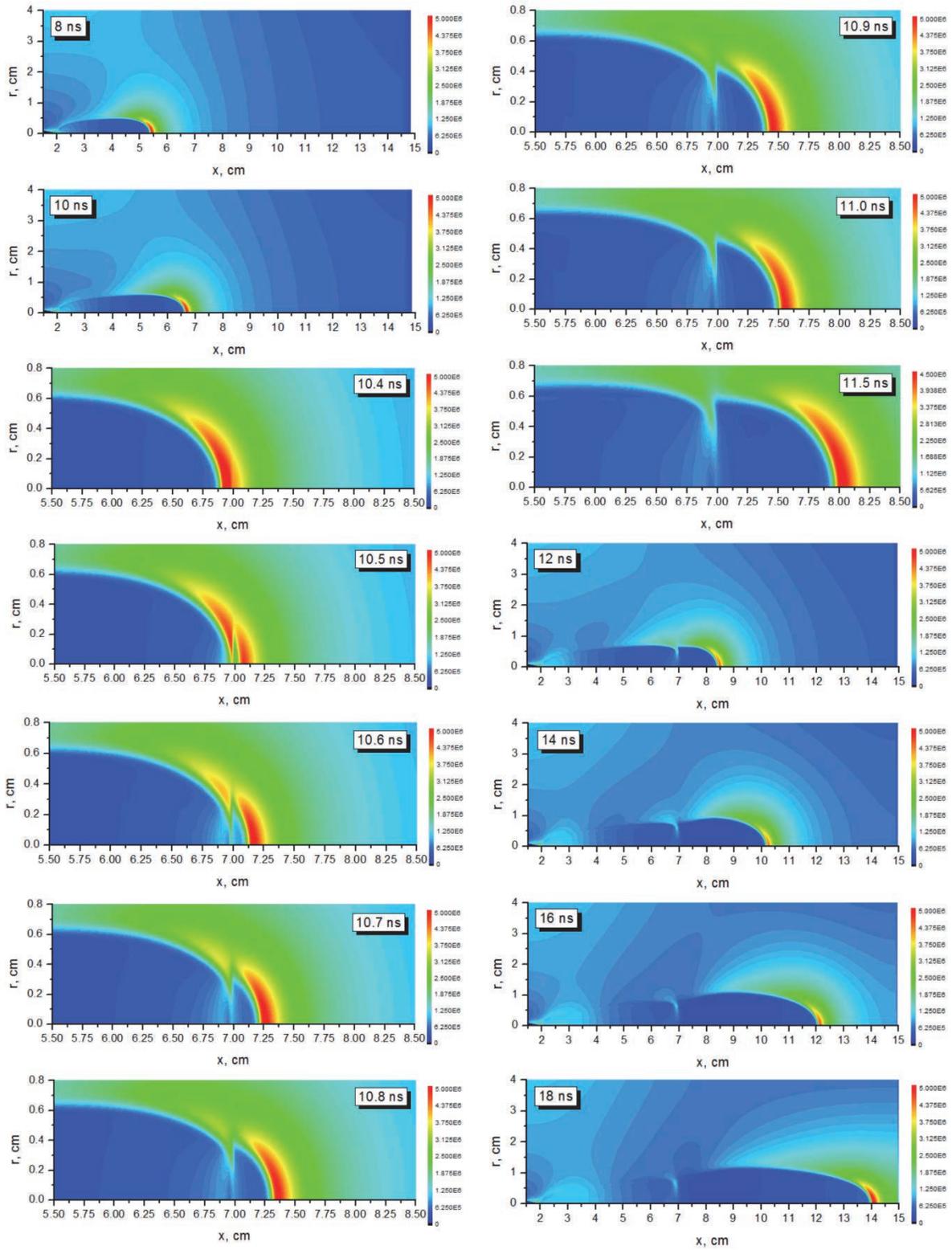

Figure 7. Contours of electric field for a streamer propagating from low- to high-density medium. $n_2/n_1 = 0.833$. $P_1 = 250$ Torr. $U = 100$ kV.

On the other hand, a similar 20% density decrease does not lead to significant new effects. Figure 7 shows the dynamics of the development of a streamer when passing the interface from a high-density medium to a low-density medium at $n_2/n_1 = 0.833$. The picture of the discharge development in this case coincides qualitatively with the case analyzed above, $n_2/n_1 = 0.483$ (Figure 3). When the streamer head reaches the density discontinuity, the discharge occurs in the low-density region, initiated by VUV photons produced by the main ionization wave and relatively weak electric fields in front of the streamer head. After that, the streamer head touches this conductive area, the charges are redistributed, the potential of the streamer head is quickly brought to the boundary of the conductive area and the streamer continues moving in the low-density gas region, gradually increasing its diameter and speed (Figure 7).

## Streamer propagation through a density gradient

If the change in gas parameters occurs at a finite length, such as, for example, in a rarefaction wave, the result of the interaction depends substantially on the length of the gradient (Figure 8). Four different cases were considered; they are a zero gradient length (discontinuity); a transition section with a length of $L = 0.1$ cm; a transition section with a length of 1 cm; and a transition section with a length of 10 cm. The calculation results demonstrate that the presence of a transition part with a length much larger than the diameter of the streamer head allows the streamer to smoothly re-shape the ionization wave form and continue moving in a medium of different density without abrupt changes in the velocity and channel diameter (Figure 8; $L = 10$ cm). Reducing the transition region to a length approximately equal to the diameter of the streamer leads to significant disturbances in the transition region (Figure 8; $L = 1$ cm). Finally, when the thickness of the transition region is noticeably smaller than the diameter of the streamer, the density gradient has almost the same effect on the development of the discharge as the gas parameter discontinuity (Figure 8; $L = 0.1$ cm, $L = 0.0$ cm).

A similar conclusion can be made when the streamer passes through a region of large change in gas density (Figure 9). When the length of the gradient is $L = 10$ cm, the streamer changes the diameter and the speed smoothly, managing to adjust to the new parameters of the medium. With a gradient length of $L = 1$ cm, the streamer shows a strong change in its parameters in the transition region. When $n_2/n_1 > 1$ (left column), the presence of a gradient allowed the streamer to form a new head and penetrate into the dense gas region, which it could not do with a discontinuous change of parameters (Figure 9). The transition to a gas of lower density with a finite gradient length is also significantly different from the case when the gas parameters change instantly (right column). In this case, the formation of a wide conducting region near the discontinuity plane is significantly smoother, and the increase in the transverse electric field in the "waist" is noticeably less pronounced.

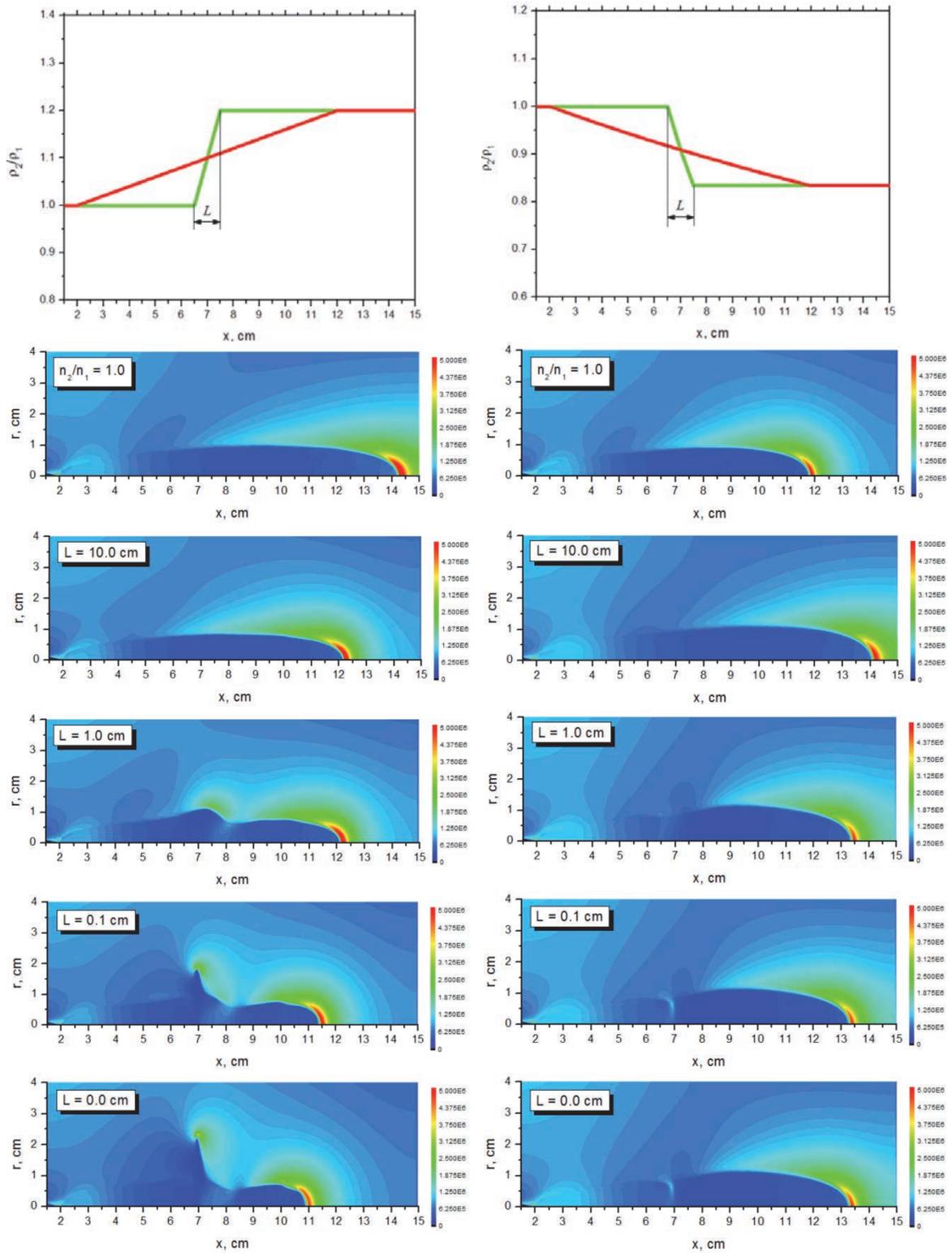

Figure 8. Contours of electric field for a streamer developing in a medium with different density gradients. $n_2/n_1 > 1$ (left); $n_2/n_1 < 1$ (right); $P_1 = 250$ Torr. $t = 19.6$ ns (left), $t = 17.1$ ns (right).

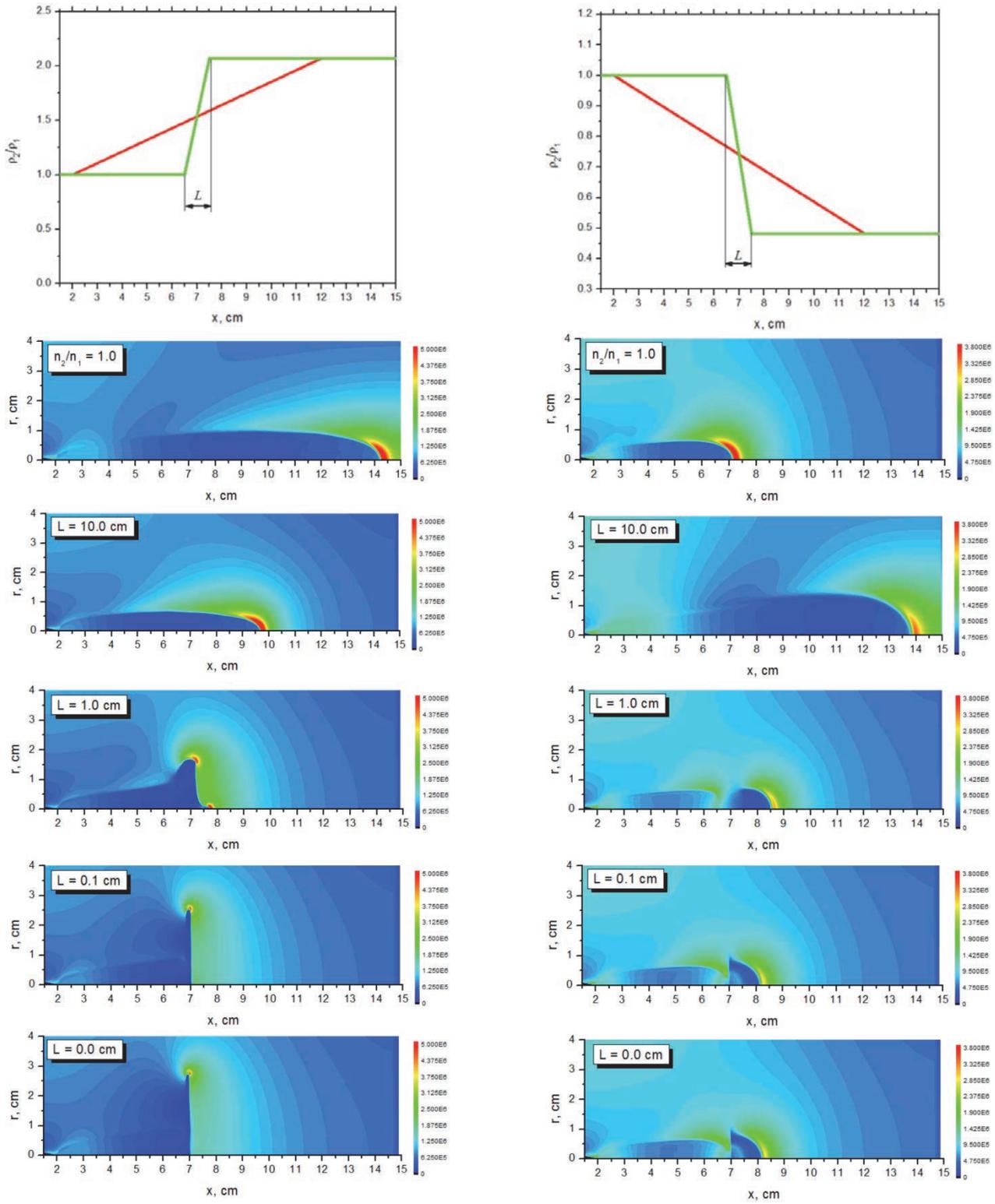

Figure 9. Contours of electric field for a streamer developing in a medium with different density gradients. $n_2/n_1 > 1$ (left); $n_2/n_1 < 1$ (right); $P_1 = 250$ Torr. $t = 19.6$ ns (left), $t = 10.8$ ns (right).

A quantitative illustration of the process of interaction of a streamer with a density discontinuity can be presented on the basis of one-dimensional parameter distributions on the axis of the streamer channel (Figure 10). When a streamer passes a gas density discontinuity of low intensity in the direction from high-density to low-density (Figure 10; row 1), both the pre-ionization of the gas by VUV radiation ahead of the main ionization wave, and the resulting plasma polarization in the electrical field of the main streamer (right column) are clearly traced. As the streamer approaches the discontinuity plane the polarization increases. When the streamer touches this region, a fast redistribution of charges occurs along the conducting region. This increases the electric field, which leads to additional ionization of the gas and a local increase in the electron concentration near the discontinuity plane (central column). The formation of a new head with a large radius leads to a decrease in the electric field maximum on the discharge axis (left column).

Qualitatively, the same picture is observed with a more significant decrease in the gas density (Figure 10; row 2). However, in this case, all processes are significantly more pronounced. The appearance of a conducting region in the low-density gas region now begins long before the streamer head approaches the discontinuity plane (right column). The plasma polarization in this region sharply increases as the main streamer approaches and peaks at time $t = 10.3$ ns. Immediately after this, as a result of the charged particles redistribution, the double layer disappears, and the streamer continues its steady-state propagation. The same process is well monitored for the electron concentration profile (central column). Ionization of the gas in the low-density region begins much earlier than in a high-density one. When the streamer approaches the formed ionization region, the electric field between two ionized regions considerably exceeds the head field before and after the discontinuity plane (left column), which leads to fast ionization of the gas in this region and to a peak in the electrons concentration.

The propagation of a streamer from a low-density gas to a high-density gas (Figure 10; row 3) is obstructed by the large initial diameter of the streamer (large diameter means relatively low electrical fields at the streamer head). In this case, the formation of a conducting "plate" on the discontinuity plane effectively shields the electric field in the central part of the discharge and significantly reduces it, despite the overall decrease in the streamer radius in a denser medium (left column). At the same time, the electron concentration in the channel decreases (central column). The right column demonstrates the process of forming a conducting "plate" and an extinction of the propagation of the charge wave when the streamer reaches the density discontinuity ($t = 10.5$ ns), and charge accumulation at the discontinuity plane ($t = 10.6$ ns). Finally the initial streamer splits into two ionization waves; one wave propagates on the surface of the discontinuity ("plate" formation) and another wave propagates along the discharge axis (formation of a streamer in a dense medium).

With a large ratio of gas densities before and after the discontinuity, the streamer cannot pass into the region containing a denser gas (Figure 10; row 4). In this case, a significant surface charge is formed on the discontinuity plane, and the ionization wave begins to propagate along the surface of the discontinuity forming a conducting "plate". In this case, the central streamer is not formed in the dense gas and the discharge development along the x-axis stops (right column). Ionization on the discontinuity surface reaches a maximum (central column), while the electric field asymptotically approaches the field of the parallel-plate capacitor. Meanwhile, the conducting "plate" develops and the influence of edge effects (left column) decreases.

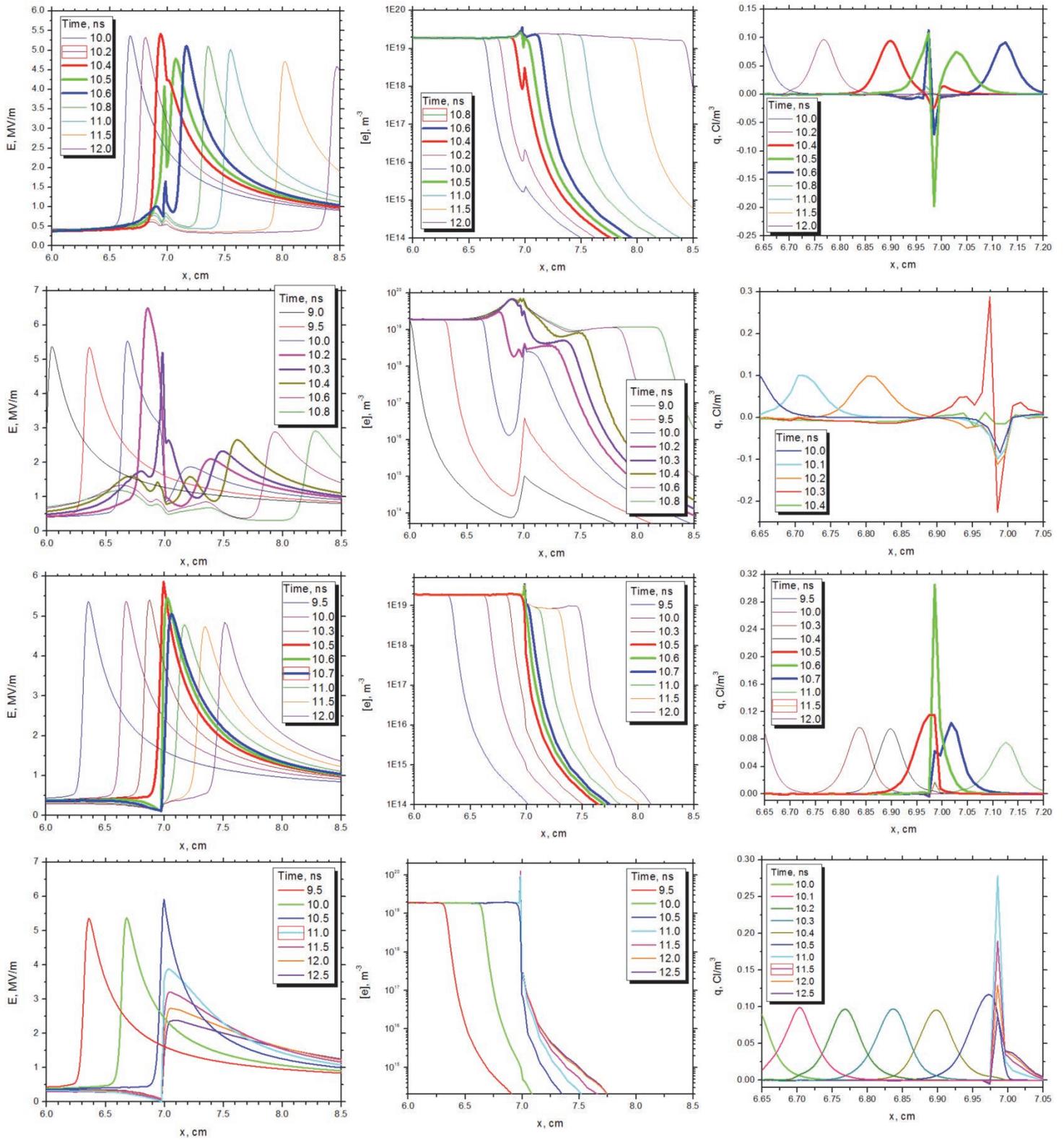

Figure 10. Dynamics of plasma parameters on the discharge axis.
Column 1: electric field; column 2: electron concentration; column 3: space charge.
Row 1: $n_2/n_1 = 0.833$; row 2: $n_2/n_1 = 0.483$; row 3: $n_2/n_1 = 1.20$; row 4: $n_2/n_1 = 2.07$.

## Discussion

The propagation of a streamer through a discontinuity in the direction of the gas density reduction leads to the expected effect, which was discussed earlier in [21]. There is a global increase in the speed of streamer propagation and an increase in the diameter of the streamer channel. In contrast, the reduced electric field ($E/n$) on the streamer head decreases. Local processes near the discontinuity surface significantly depend on the pre-ionization of the low-density gas by VUV photons emitted at the ionization wave front. In the region of the pre-ionized low-density gas, even before its contact with the main streamer body, an electrodeless discharge develops. The conductive region created in this way leads to an abrupt change in the parameters of the streamer on the dense gas/rarefied gas interface. However, in general, the development of the streamer continues and the conductive channel passes through such a discontinuity almost freely.

A completely different result is observed for the development of a streamer across the rarefied gas/dense gas interface. In this case, when the critical value of the density discontinuity is reached, the streamer is completely stopped, and its further development occurs along the discontinuity plane rather than along the discharge axis. In order to discuss in more detail the mechanism of this process, let us illustrate schematically the propagation of the ionization wave near the density discontinuity (Figure 11). This figure shows the simplified shape of the streamer head when it is passing the interface with a denser medium with a density ratio $n_2/n_1 = 1.2$. In this case the shape of the ionization wave can be approximated by a cylinder with a hemispherical head (Figure 11, below).

When a streamer propagates from a less dense medium to a denser one, the ionization wave slows down near the streamer axis, since this part of the wave starts to interact with the denser medium first. For further analysis, we assume that the ionization wave propagating into a denser gas, nevertheless, retains a hemispherical shape. In this case, the unperturbed ionization wave at the head of the streamer has a curvature radius $R_1$, and the disturbed one has a radius $R_2 > R_1$.

One can assume that the loss of stability of the original ionization wave, the formation of a conductive "plate" and finally, the streamer stop will occur if the reduced electrical field $E/n$ on the side of the surface of the "undisturbed" streamer becomes higher than the value of $E/n$ on the axis (in the region where the ionization wave propagates into the high-density gas in Figure 11, above). Then the radial speed of the ionization wave propagation becomes faster than the speed of its development along the axis, and a quasi-surface discharge propagating along the surface of the density discontinuity is formed.

Obviously, the maximum field outside the axis of the streamer will be at the intersection of the unperturbed ionization wave with a discontinuity (point "A" in Figure 11, below). Using the assumptions discussed above the geometry of the streamer head at this point can easily be estimated using the following simple geometric relations:

$$R_2 \sin(\alpha_2) = R_1 \sin(\alpha_1)$$

$$R_1 \cos(\alpha_1) + v_1 t = R_1 \quad (1)$$

$$R_2 \cos(\alpha_2) + v_2 t = R_2$$

Here, the angles $\alpha_1$ and $\alpha_2$ characterize the streamer penetration into a dense medium, $v_1$ and $v_2$, respectively, represent the speed of propagation of the streamer head on the axis in both, the low- and high-density environment. The solution of system (1) for known values of $R_1$, $\alpha_1$ and $n_2/n_1$ gives the values of $R_2$ and $\alpha_2$. Consequently, the reduced electric field at different points of the streamer head penetrating into a denser medium could be derived.

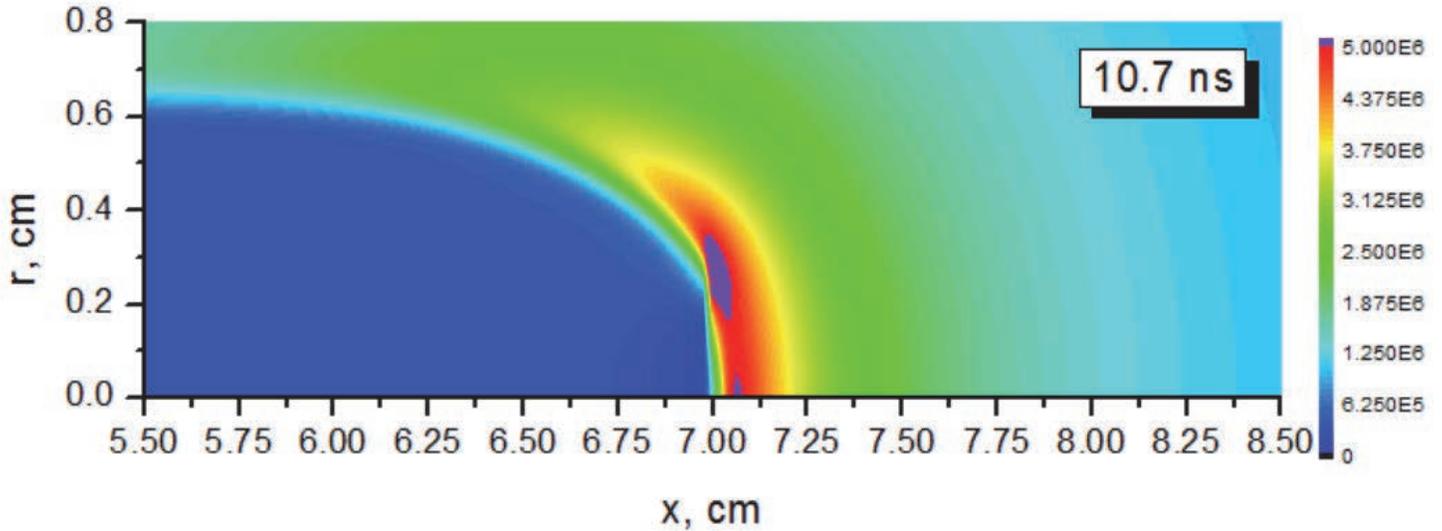

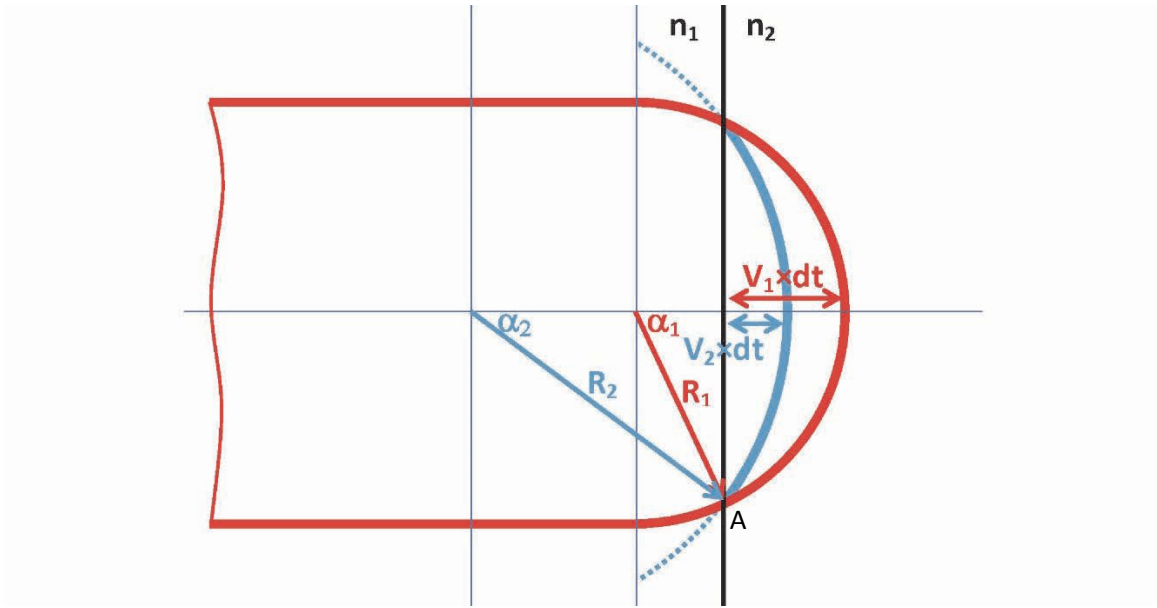

Figure 11. Diagram of a streamer propagation through a density discontinuity at $n_2/n_1 > 1$. Above: the shape of the streamer head interacting with the interface with a denser medium according to the results of 2D numerical simulation. $n_2/n_1 = 1.2$. $P_1 = 250$ Torr. $U = 100$ kV. Bottom: approximation of the shape of the streamer head (the position of the density discontinuity is marked by the black line).

One can use the analytical model of Baselyan-Raizer [1], which gives a relation between the radius of curvature of the ionization wave at the streamer head, $R$, and its velocity v:

$$v = \frac{\gamma_{im} R}{(2k_f - 1)\ln(n_m/n_0)} \quad (2)$$

where $\gamma_{im}$ is the frequency of gas ionization at the point where the field reaches the maximum $E_m$. $\gamma_{im}$ depends only on the type of gas and the value of the local electric field $E_m/n$. $k_f$ is the coefficient in the formula approximating the dependence of the frequency of ionization on the local electric field near the field maximum:

$$\gamma_i = \gamma_{im}(E/E_m)^{k_f}, \quad (3)$$

$n_m$ is the electron concentration at the point of maximum field and $n_0$ is the concentration of background electrons. The field on the symmetry axis of a cylindrical channel ($\alpha_1 = 0$) with a hemispherical head can be approximately estimated as [1]:

$$E(0) \approx \frac{1}{2} E_{sphere} = \frac{Q}{2\pi\varepsilon_0 R_m^2} \quad (4)$$

Within the framework of the adopted approximations $E_m \sim E(0)$. The dependence of $\gamma_{im}$ on $E_m$ was calculated using the two-term approximation of the Boltzmann equation using the same method as for the numerical simulation. Approximation (3) of the dependence of the ionization frequency on the field aptly describes the numerical solution of the Boltzmann equation with $k_f = 2.5$ (Figure 12). This allowed the analytical integration of the electrons avalanche development from the origination point to the point of its merging with the streamer head [1].

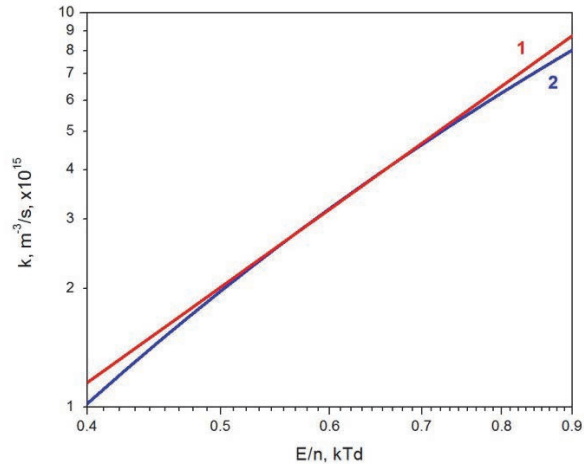

Figure 12. Dependence of the ionization frequency on the local field. 1 - analytical approximation (4) with $k_f = 2.5$; $E_m = 600$ Td. 2 - numerical solution of the Boltzmann equation.

The distribution of the field with the distance from the axis of symmetry for the model head shape (Figure 11) can be described by an approximated formula:

$$E(\alpha_1) \approx E(0)\left(1 - \left(\frac{1}{2}\sin(\alpha_1)\right)^2\right) \quad (5)$$

Approximation (5) is obtained from the numerical simulation of the distribution of the electrostatic field around an equipotential cylindrical surface with a hemispherical curve (Figure 13).

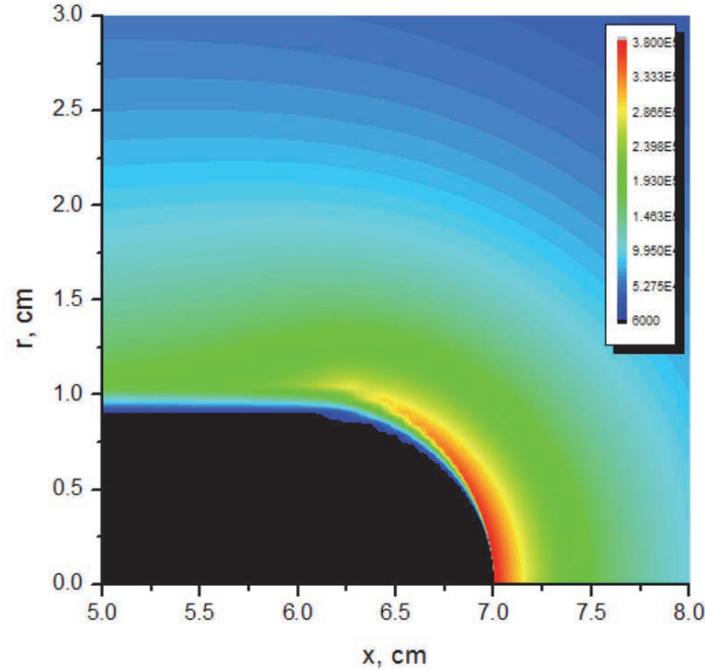

Figure 13. The distribution of the electric field for the model shape of the streamer head.

Equations (1-5) make it possible to obtain the dependence to $R_2(\alpha_1)$ for each fixed initial streamer radius $R_1$ at a known streamer voltage. Then, for each value of $\alpha_1$, which determines the penetration of the streamer into the high-density gas region, one can calculate the magnitude of the electric field (equation 5) at the point of the off-axis maximum (point A in Figure 11).The field on the axis of symmetry of the new wave can be calculated by the radius of the curvature of the ionization wave in a high-density gas, $R_2$ (equation 3).

Now it is easy to find the stability limits of the initial ionization wave versus the magnitude of the gas density jump. The streamer is stable when the ratio of the reduced fields on the lateral surface $E_1/n_1$ ($\alpha_1 = 90^0$) and on the axis of symmetry $E_2/n_2$ is less than one:

$$\frac{E_1/n_1}{E_2/n_2} < 1 \qquad (6)$$

In the opposite case, the ionization rate on the lateral surface of the streamer exceeds the ionization rate on the axis, and the development of the streamer along the surface of the density discontinuity becomes dominant. Figure 14 shows the calculation results for the model (1-6). It should be noted that the model (1-6) does not determine the initial radius of the streamer and its head. Moreover, the radius of the head curvature and the channel radius may differ significantly (Figure 9). Therefore, Figure 14 shows the range of parameters from the minimum radius of curvature on the head of the streamer ($R_1$(head) ~ 7 mm at $U$ = 100 kV) to the radius of the channel ($R_1$(channel) ~ 10 mm). Estimates of the streamer radii at 50 and 200 kV were made using an analytical model [26], which gives a relationship between the streamer radius and the frequency of ionization and photoionization in gas.

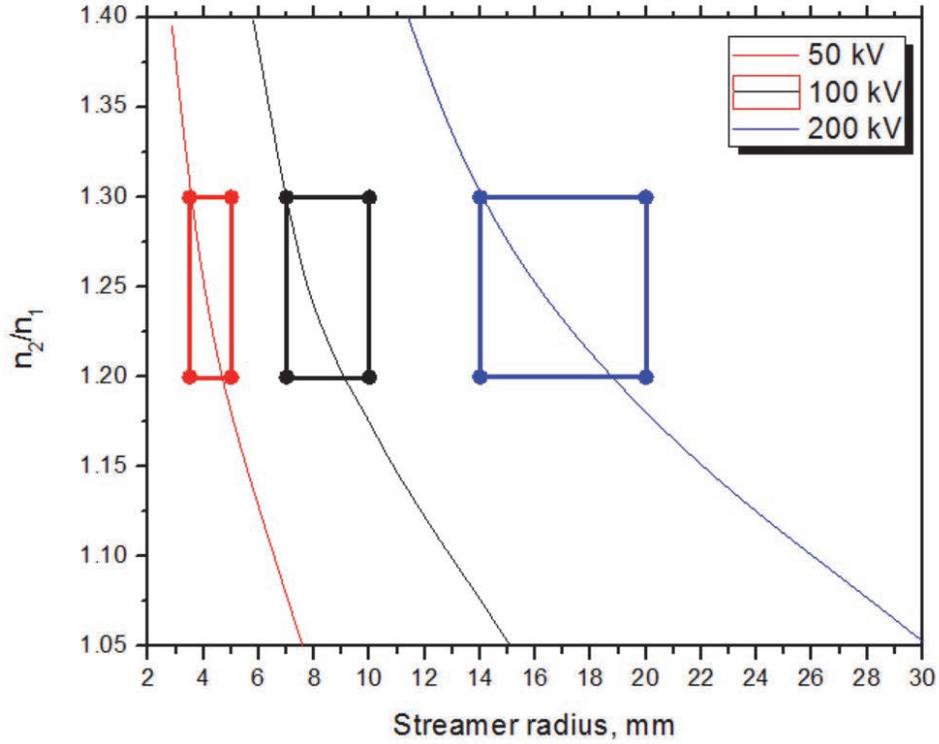

Figure 14. Critical density ratio $n_2/n_1$ estimation using the model (1-6) depending on the radius of the streamer channel. Streamer head potential was 50, 100 and 200 kV, air pressure 250 Torr.

According to model (1-6), the critical intensity of the discontinuity in gas density through which a streamer can or cannot penetrate is $n_2/n_1 \sim 1.2$-$1.3$ in the voltage range of 50 to 200 kV. This result is in good agreement with the results of the numerical calculations (Figure 5). The model calculations show that the ability of the streamer to penetrate through the discontinuity of the gas density decreases with an increase in its radius, and increases with voltage increase. This behavior is intuitively obvious because an increase in the field at the streamer head makes it more resistant to external perturbations.

Note that the assumptions made in this analysis do not allow for any kind of rigorous quantitative analysis of the stability ranges of streamer development, but they do allow for the identification of the main mechanisms responsible for the development of instability. The mechanism which controls the streamer propagation through the density discontinuity is a sharp decrease in the frequency of ionization due to an increase in the density of the gas and the decrease in the magnitude of the reduced electric field $E/n$ caused by it. Equation (2) demonstrates a linear dependence of the ionization wave velocity on the gas ionization frequency, which signifies an exponential dependence on the magnitude of the reduced field, $E/n$. A decrease in speed of the ionization wave in a dense medium immediately leads to an increase in the radius of curvature of the ionization wave front near the discharge axis, and to the formation of an off-axis field maximum. The appearance of this secondary maximum immediately leads to a loss in stability of the initial ionization wave.

# Conclusions

Modeling of the streamers development in a highly inhomogeneous gas showed that the discharge characteristics change dramatically when the streamer reaches the boundary between the areas with different gas densities. In the case of streamer propagation from a high-density gas region to a low-density gas region, in the low-density region the diameter of the streamer channel increases, while the peak electric field in the streamer head and the electron density in the streamer channel decrease. Ionization of the gas in the low-density region starts before the streamer head reaches the boundary between regions with different gas densities. This plasma is generated in the low-density region due to the photoionization of gas molecules by VUV photons emitted at the head of the streamer, when it is still located in the region of high gas density. As a result, the first ionization wave is formed in the low-density region. This wave is followed by a second ionization wave formed at the moment when the streamer head reaches the gas density discontinuity plane. The second ionization wave attenuates quickly, but provides an increase in the channel conductivity in the region where its diameter is at a minimum (at the transition point from the region of high gas density to the region of low gas density).

The development of a streamer during its propagation from a region of low gas density to a region of high gas density demonstrates qualitatively different properties. With a small density difference, the streamer goes into the higher density gas, reducing its diameter, speed, and at the same time increasing the field at the head and the electron concentration in the channel. When the density increase exceeds $n_2/n_1 = 1.2$, the interaction leads to a sharp decrease in the ionization rate and the velocity of the streamer propagation along the axis. Simultaneously, there is a decrease in the radius of the curvature of the surface of the ionization wave located closer to the lateral surface of the streamer channel. As a result, the off-axis maximum of the electric field is observed, and the ionization wave forming the streamer head loses its stability. Further development of the streamer generates a conductive "plate" located in the plane of discontinuity of the gas density. Thus, with a sharp increase in gas density, putting it above the critical value, the streamer can no longer overcome such a discontinuity. The critical value of the density jump is determined by the ratio of fields on the axis of the streamer head, on its side surface and, to a lesser extent, on the characteristics of nonlocal gas pre-ionization by high-energy photons emitted at the ionization wave front. In air, this critical value is close to 1.2, which means that the streamer cannot overcome the discontinuity of densities created (for example, by a direct shock wave with a Mach number greater than $M = 1.15$). This conclusion demonstrates the fundamental limitations of the possibility of creating nonequilibrium plasma in supersonic and hypersonic flows, where there are numerous intense shock waves. In such flows, in order to create nonequilibrium excitation, it is necessary to organize the development of a discharge from a region with a higher density in a region of lower densities.

Thus, we found conditions when the gaseous medium for short times began to possess unidirectional conductivity. Here, a discontinuity in the gas density formed a kind of "gas-dynamic diode" that allowed the plasma channel to propagate in one direction and blocked its development in another. It is worth noting that, instead of using the density discontinuity to create the effect of a directionally conducting medium, the interface between gases with different ionization cross sections could be used.

The simulations show the possibility of a streamer passing through the region of higher gas density in the case of a finite gradient value; such as, for example, in a rarefaction wave or mixing layer in the case of different gases. Increasing the length of the gas density gradient (the width of the fan of compression/rarefaction waves) allows the streamer to overcome this gradient even if the density ratio is sufficiently high in the initial and final states of the gas. The effect becomes decisive when the characteristic length of density change becomes comparable with the radius of the streamer head.

## Acknowledgements

This work was supported by the Russian Science Foundation (project No 17-12-01051) and MetroLaser/NASA project "Short Pulsed Laser Techniques for Measurement of Multiple Properties in High Enthalpy Facilities".

# Additional Information
## Competing interests
The authors declare no competing interests.